\documentclass[fleqn,10pt]{wlscirep}
\pdfoutput=1 
\usepackage[utf8]{inputenc}
\usepackage[T1]{fontenc}
\usepackage{color, colortbl}
\usepackage{threeparttable}
\DeclareUnicodeCharacter{394}{$\alpha$}
\title{Nano-size nature of the $\alpha$-FAPbI$_3$ by means of large-scale ab initio simulations}

\author[1,+]{Virginia Carnevali}
\author[1,+]{Lorenzo Agosta}
\author[1]{Vladislav Slama}
\author[1]{Nikoalos Lempesis}
\author[1]{Andrea Vezzosi}
\author[1,*]{Ursula Rothlisberger}
\affil[1]{Laboratory of Computational Chemistry and Biochemistry,Institute of Chemical Sciences and Engineering, Swiss Federal Institute of Technology (EPFL), Lausanne, Switzerland}

\affil[+]{these authors contributed equally to this work}

\affil[*]{ursula.roethlisberger@epfl.ch}

\begin{abstract}
Formamidinium-lead-iodide (FAPbI$_3$) has established itself as the state of the art for high solar-energy conversion efficiency in perovskite-based solar cells. At room temperature, FAPbI$_3$ has a peculiar crystal structure with tetragonal symmetry where the PbI$_6$ framework is distorted from the perfect cubic structure, while the FA molecules are randomly rotated. This is  well known experimentally, but the theory is still deficient in describing FAPbI$_3$ with appropriate models in which the system size is adequately taken into account. Using ab initio molecular dynamics at 300 K and first-principle calculations, we prove that, in order to obtain a proper description of the system, three factors must be satisfied simultaneously: the band gap, the minimization of structural distortion, and the zeroing out of the dipole moment. We show that the net dipole moment zeroes out as the system size increases due to PbI$_6$ octahedra distortions rather than FA rotations. We also show that the band gap oscillations in temperature are correlated to octahedra tilting. The optimum between simulations and experimental properties indicates that FAPbI$_3$ is properly described by a system size approaching the nano-scale.    
\end{abstract}

\begin{document}
\definecolor{Gray}{gray}{0.9}
\sloppy

\flushbottom
\maketitle
%
%
\thispagestyle{empty}

\section*{Introduction}
Modern photovoltaics (PV) point to metal halide perovskites (MHP) as one of the most promising classes for active layer (AL) materials.\cite{zhang_high-efficiency_2017,jeong_pseudo-halide_2021} Thanks to their remarkable optical properties such as a high absorption coefficient, tunable bandgap, high charge carrier mobility, and low exciton binding energy, perovskite-based solar cells (PSCs) lead to high solar cell performance (PCE). Perovskites have the chemical formula ABX$_3$, where A is an organic or inorganic cation, B is a divalent metal cation, and X is an anion of the halogen group. Efficient and stable PSCs require uniform and defect-less perovskite thin films at large areas, which can improve charge transport, suppress non-radiative energy loss, and minimize device degradation pathways.\cite{liu_holistic_2020,cao_quantum_2019} So far, among all MHPs, formamidinum-lead-iodide (FAPbI$_3$) in the cubic symmetry ($\alpha$-FAPbI$_3$) gives the best PV performance. FAPbI$_3$ reported experimental band gaps are in the range of 1.45-1.51 eV,\cite{min_efficient_2019,li_stabilizing_2016,eperon_formamidinium_2014,jeong_pseudo-halide_2021} which is close to the ideal single-junction Shockley-Queisser band gap of 1.31 eV.\cite{shockley_detailed_1961} Besides, compared with other perovskites such as methylammonium-PbI$_3$ (MAPbI$_3$), FAPbI$_3$ exhibits a long-term thermal stability due to higher activation energies for thermal degradation.\cite{chen_advances_2021} Hence, the narrow band gap and better thermal stability establish FAPbI$_3$ as the most promising AL among the perovskites for single-junction PSCs with PCE over 30\%.\cite{noauthor_best_nodate}

However, further improvements in the stability and performance of FAPbI$_3$ have prompted experimentalists to seek theoretical support to understand the pivotal mechanisms involved in surface passivation and charge transport, which are critical to guide a new design of PSCs. \cite{han_tautomeric_2023,li_buried-interface_2024,park_low-loss_2023,jeong_carbazole_nodate} This requirement has challenged theoreticians to design surface and bulk models of FAPbI$_3$ with an appropriate level of accuracy that can mimic the main electronic and structural features of FAPbI$_3$ films. These include phenomena such as homogeneous crystallization from solution, halide segregation, and defect formation, that still need to be resolved at the atomistic level. These collective mechanisms require simulations with an adequate level of precision while allowing the number of atoms to be greatly increased. Because of the system size requirements, the type of simulation techniques used to simulate FAPbI$_3$ phases have been in the framework of classical molecular dynamics (CMD),\cite{ahlawat_combined_2021,ahlawat_atomistic_2020} density functional tight binding approach,\cite{vicent-luna_efficient_2021,uratani_simulating_2020} or monte carlo methods.\cite{shu_monte-carlo_2018,labrim_magnetic_2015} In this regard, also machine-learning force fields trained on ab initio molecular dynamics (AIMD) data are really promising solutions because they allow CMD to be performed with the AIMD accuracy, provided the physics is correctly described in the training data set.\cite{jinnouchi_phase_2019,liang_toward_2024} It is evident that a correct description of FAPbI$_3$ at the level of quantum mechanics is of paramount importance. 

In this context, there is an important amount of computational literature on FAPbI$_3$ - and MHPs in general -  where different periodic models and levels of theory are used, sometimes leading to contradictory results for fairly fundamental properties such as band structure and band gap,\cite{liu_role_2021,wang_structural_2020,zhao_recent_2022} suggesting a lack of predictive ability. 
Several computational methods have been adopted to obtain the experimental band gap of FAPbI$_3$, spanning from density functional theory (DFT), to hybrid functionals with different parameter schemes, to the GW approximation of many-body perturbation theory, or a combination of them. In addition, spin-orbit coupling (SOC) is usually implemented in presence of Pb/Sn atoms, which leads to a significant lowering of the band gap.  A fairly comprehensive summary of the band gap calculated for FAPbI$_3$, depending on the level of theory used, is provided in this review\cite{raeisianasl_review_2022}. The band gap can be also influenced by the choice of the system size. The mismatch between computed FAPbI$_3$ properties with different level of theories might be due to the choice of small system sizes, which force an ordered molecular orientation of ferroelectric multi-domains, which is not observed experimentally.\cite{amat_cation-induced_2014,bokdam_role_2016,jono_theoretical_2019,ummadisingu_crystal-size-induced_2021} There are a few examples in literature which adopt larger systems size, obtaining interesting results. Ma et al.\cite{ma_nanoscale_2015} showed that to correctly capture the electronic structure of MAPbI$_3$ it is necessary to treat the system at the nano-scale.

In this letter, we want to address the key questions: what are the system size and level of theory needed to capture the structural and electronic properties of FAPbI$_3$ and, is the ground state description sufficient or simulations at finite temperature are required? These are crucial questions from both theoretical and experimental perspectives.

Using IMD at 300 K and first-principle calculations, we characterized the structural and electronic properties of  $\alpha$-FAPbI$_3$ with increasing size of the simulation cell, both with PBE and PBE0 methods. SOC was also considered when allowed by memory and/or computational resources. Our simulations show that it is necessary to use sufficiently large simulation cells (at least a 768-atom cell) in which all degrees of freedom (atomic positions and simulation cell) are relaxed in order to obtain an non-distorted FAPbI$_3$ structure description, a good band gap and a small residual electron dipole. At 0 K, to achieve the true energy minimum, it is also essential to set up an initial system configuration in which the FAs are randomly oriented according to 3-fold symmetry. Furthermore, with large simulation cells (from a 2592-atom cell upwards), the PBE is already able to describe the electronic band gap when calculated as the average of several finite-temperature equilibrated AIMD snapshots. We also show that the net dipole moment of the system goes to zero as the cell size increases and this is related to long-range effects due to the PbI$_6$ octahedra tilting. Finally, we demonstrate that band gap oscillations in temperature are related to the octahedra tilting.

\section*{Results}
\subsection*{Characterization of $\alpha$-FAPbI$_3$ at 0 K}
The calculated band gaps are reported in Tab.\ref{tbl:bg}. Band gaps were computed for the usual primitive cell (12-atom) and for larger simulation cells with an even number of primitive cells. The supercell choice is intended to reflect the importance of not breaking the octahedra tilting pattern,\cite{woodward_octahedral_1997} while the 12-atom cell to have our own internal reference. At 0 K, we ran two series of simulations using as the initial structure the $\alpha$-FAPbI$_3$ from the material project database.\cite{jain_commentary_2013} In the first case (relax), only the atomic positions were optimized keeping the simulation cell cubic, while in the second (vc-relax), both the atomic positions and the lattice parameters were optimized. We also set up two initial configurations: $(i)$ FA molecules are kept aligned, in order to have a meaningful comparison between the 12-atom cell and all the other supercells; $(ii)$ FAs are pseudo-randomly oriented (from the 96-atom cell upwards), where the overall orientation of FAs preserves the 3-fold symmetry and minimizes the dipole moment of the system (Supplementary Fig.1). K-point sampling was increased until convergence was achieved. 

For both relax and vc-relax simulations, the band gap converges with the increase of the k-point grid but at different values, indicating that the relaxation of all the cell degrees of freedom is necessary. For all the simulation cells with aligned FAs the relax band gap converges around 1.48 eV. At first glance, this is a good value for the $\alpha$-FAPbI$_3$ band gap and can be justified by the fact that, by keeping the simulation cell cubic, we are, in a sense, simulating the average $\alpha$-FAPbI$_3$; however, with a closer look at the structure, we show that this is a fictitious artefact. Indeed, in addition to the correct band gap, it is necessary to verify that the cell distortions from the $\alpha$-FAPbI$_3$ structure upon vc-relax and the total dipole moment of the supercell have also physical meaning. The vc-relax band gap of the 96-atom cell converges to slightly higher value respect to the 12-atom one.
This is related to the significant distortion of the crystal structure from the cubic symmetry after vc-relax, increasing the overlap of the electronic orbitals and consequently opening the band gap.\cite{meloni_valence_2016} To estimate the structural distortions of the crystal, we have computed the mean squared error (MSE) between the perfect $\alpha$-FAPbI$_3$ structure and the vc-relax one as well as the distribution of the octahedra tilting angles  after vc-relax for the two cases - with and without pseudo-random FA orientation (Tab.\ref{tbl:DistDip}). The pseudo-random orientation of the FAs in the starting configuration allows cubic symmetry to be maintained almost perfectly, as the MSE is reduced by two orders of magnitude for all supercells compared to the case with fully aligned FAs. 
The distributions of the octahedra tilting angles of the structures optimized from the pseudo-random FA configurations average 0 degrees as they should (Supplementary Fig.2) and their amplitudes converge with increasing system size (Tab.\ref{tbl:DistDip}). The 96-atom cell has a wide spread distribution because of the significant lattice distortions after vc-relax. In contrast, the structures optimized from all-aligned FAs do not present the typical octahedra tilting pattern. Local distortions of the octahedra due to a collective upward or downward motion of I ions compensate for the strongly directional dipole due to the aligned FAs (Supplementary Fig.3). The vc-relax band gap is only improved for the 2592-atom cell and may be related to the fact that only this supercell allows full relaxation of all degrees of freedom because both the 3-fold symmetry for FAs and the octahedra tilting pattern are satisfied. For all the supercells, the potential energy of the pseudo-random case is always lower than that of the all-aligned case, which means that the pseudo-random FA orientation is a more convenient choice for the system.  

By adding SOC, the band gaps of the 12- and 96-atom cells converge to slightly different values - 1.55 eV and 1.60 eV - again indicating a potential problem in the description of the electronic structure related to the system size. Remarkable is what is obtained with the PBE0 functional. It is known that for elements such as Pb, SOC and PBE0 contributions should cancel each other out.\cite{borlido_large-scale_2019} From our 0 K simulations, this effect only comes into play at the convergence of the k-point for the 96-atom cell, whereas it is  absent for the 12-atom cell; by moving from the 12-atom cell to the 96-atom cell, the electron charge localization decreases as the band gap correction due to the inclusion of PBE0 decreases by 0.65 eV, allowing for SOC compensation and converging to a band gap to 1.60 eV. It is evident that the size of the system plays a crucial role for FAPbI$_3$ both in order to obtain the right description of the electronic structure by cancelling out the SOC-PBE0 contributions and in order to ensure the structural properties of the system converge well. 

\subsection*{Characterization of $\alpha$-FAPbI$_3$ at 300 K}
\begin{figure}[ht!]
  \includegraphics[width=1.0\textwidth]{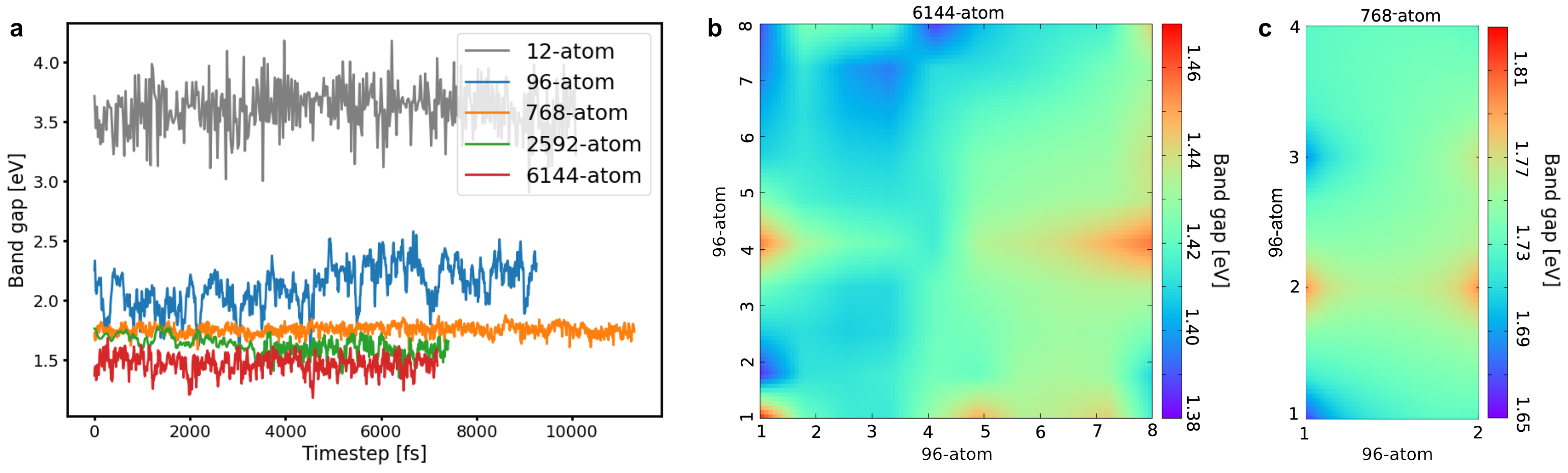}
  \caption{(a) Time variation of the band gap along the NPT-F trajectory at 300 K for different supercells. Spatial band gap variation on 64 96-atom cells contained in a 6144-atom FAPbI$_3$ cell (b) and on 8 96-atom cells contained in a 768-atom cell (c).}
  \label{fgr:LBG}
\end{figure}

The FAPbI$_3$ band gap was also calculated at 300 K by performing flexible NPT AIMD (NPT-F) with PBE and PBE0 level of theory for a minimum of 7 ps up to 11 ps (Tab.\ref{tbl:bg}). The band gap was computed as the average along the equilibrated AIMD trajectory (Fig.\ref{fgr:LBG}-a). We run in NPT-F to avoid any kind of symmetry restriction to the system. Since we are operating at 300 K, the initial orientation of the FAs is no longer important, as the kinetic energy is sufficient to randomize the FA orientation after a few AIMD steps. We calculated a band gap in temperature for the 6144-atom cell of 1.47 $\pm$ 0.08 eV that matches very well, within error, to that measured experimentally, whereas for all simulation cells, PBE0 always overestimates the band gap. However, we note that by adding the converged SOC contribution of 0.6 eV estimated from the 96-atom cell at 0 K, we obtain the correct band gap value even for the 6144-atom cell with PBE0, indicating that the system size is ideal for reproducing the correct electronic state. 

\begin{figure}[ht!]
  \includegraphics[scale=1.3]{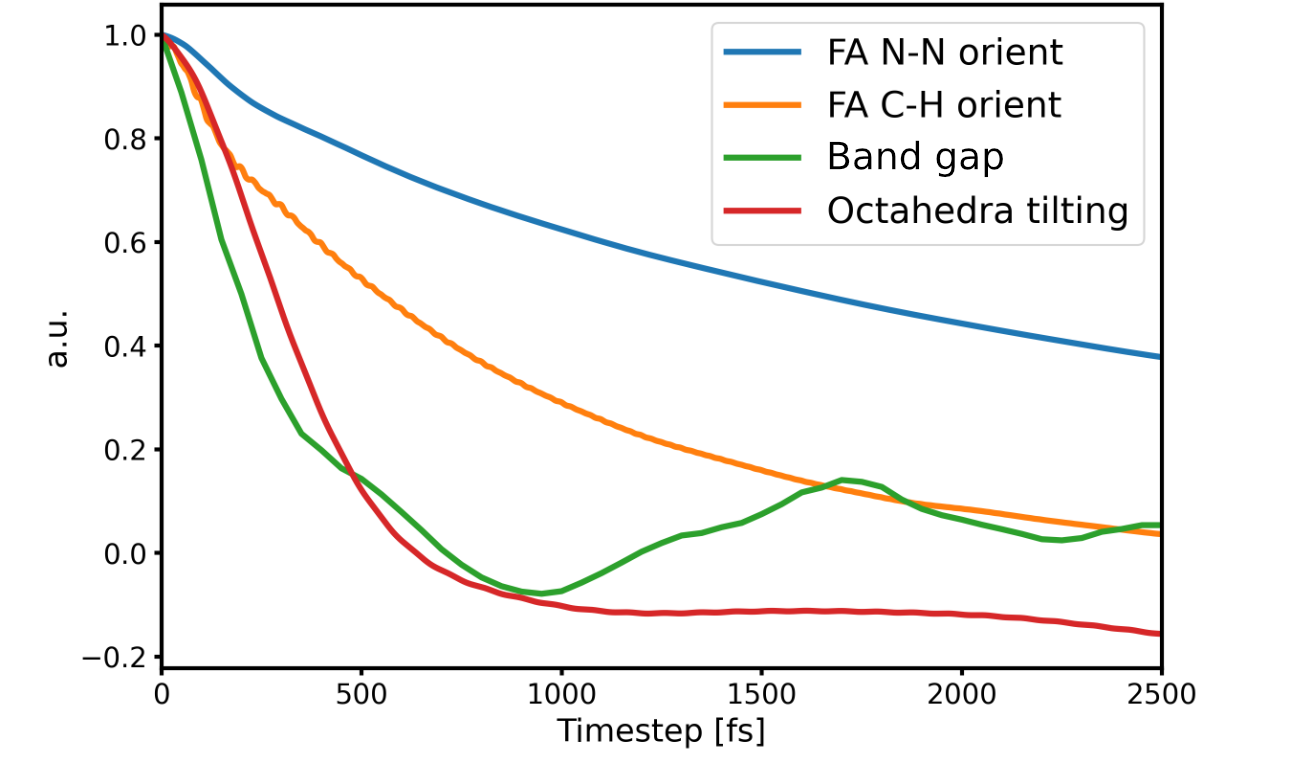}
  \caption{ Time correlation function characterizing the rotational dynamic of FA along the N-N (blue curve) and C-H (orange curve) axes, as well as the octahedra tilting (red curve), and band gap oscillations (green curve). The analysis was done on the 6144-atom cell.}
  \label{fgr:CorrFunc}
\end{figure}

We also exploit the larger systems size allowed by AIMD together with its statistical approach to study the in time band gap dependence from the local lattice distortions. We have divided the 6144-atom and 768-atom cells into 64 and 8 96-atom cells, respectively, and computed the local band gap by projecting the density of states locally along the NPT-F run ( Fig.\ref{fgr:LBG}-b,\ref{fgr:LBG}-c). The 3D disposition of the supercells was flattened in a 2D map for better visualization (Supplementary Fig.4). There is a local variation in band gap for both supercells of about 1 eV, but in the 6144-atom cell there are more defined band gap domains, while for the 768-atom cell the band gap is more homogeneous throughout the supercell. This again indicates the importance of the size of the system in $\alpha$-FAPbI$_3$, in order to detect the formation of the band gap domains. It further indicates that this is a phenomenon that has to be related to the process of long-range relaxation, stressing once again the importance of correctly describing the mechanisms that we have previously shown to be related to the size of the system, such as the octahedra tilting. To quantify the connection between band gap fluctuations and octahedra tilting, we have calculated the time correlation function of the band gap oscillations and octahedra tilting for the 6144-atom cell (Fig.\ref{fgr:CorrFunc}), getting a correlation time of about 25 fs and 30 fs, respectively. The two correlation times are in very good agreement. FA has two characteristic rotational correlation times of 80 fs and 250 fs for the C-H and N-N vectors respectively (Fig.\ref{fgr:CorrFunc}), that are far from the value calculated for the band gap oscillations.
We have also computed the time correlation functions of the bottom of the conduction band and top of the valence band eigenvalues obtaining a trend comparable with that of the band gap (Supplementary Fig.5). Local variations in the band gap might indicate that FAPbI$_3$ can potentially absorb photons with different wavelengths in different regions of the sample, which may be a further explanation of why this material performs as well as AL. The electronic charge distribution due to crystal motions in halide perovskites have implications on thermal fluctuations in the electronic structure. It has been shown that off-centering of Sn$^{2+}$ and Br$^-$ motions widens the band gap of CsSnBr$_3$.\cite{fabini_dynamic_2016} Additionally, the top of the valence band and the bottom of the conduction band of FAPbI$_3$ are dominated by I$^-$ and Pb$^{2+}$ contributions, respectively.\cite{wang_structural_2020} For these reasons, we can expect that the fluctuations in the band gap occur on timescales similar to those characterizing the octahedra tilting fluctuations. Having the right system size is again crucial to correctly describe the octahedra tilting pattern. The last point needed to verify the proper description of FAPbI$_3$ structure is the dipole moment.

\begin{table}[ht!]
  \caption{FAPbI$_3$ band gap for different system sizes at several theory levels and system optimization schemes. The initial configurations for the 0 K simulations were chosen with all FAs aligned except the one indicated with a star superscript. AIMD was performed at 300 K in NPT-F and the band gaps were computed as averages along the trajectories after equilibration. Values in brackets are estimated from SOC and PBE0 calculations because they require too much memory for accessible resources.}
  \label{tbl:bg}
  \centering
  \resizebox{\textwidth}{!}{\begin{tabular}{c|ccccccc|c}
    \hline
    \rowcolor{Gray}
     \begin{tabular}[c]{@{}c@{}}Simulation\\ cell \end{tabular} &  \multicolumn{1}{c}{\begin{tabular}[c]{@{}c@{}}NPT-F\\PBE\\ 300 K\\ 
     {(}eV{)}\end{tabular}} & \multicolumn{1}{c}{\begin{tabular}[c]{@{}c@{}}NPT-F \\ PBE0\\ 300 K\\ {(}eV{)}\end{tabular}}& \multicolumn{1}{c}{\begin{tabular}[c]{@{}c@{}}relax \\PBE\\ 0 K \\ {(}eV{)} \end{tabular}}  & \multicolumn{1}{c}{\begin{tabular}[c]{@{}c@{}}vc-relax \\PBE\\ 0 K \\ {(}eV{)}\end{tabular}}& \multicolumn{1}{c}{\begin{tabular}[c]{@{}c@{}}vc-relax \\ PBE+SOC \\ 0 K \\ {(}eV{)}\end{tabular}} & \multicolumn{1}{c}{\begin{tabular}[c]{@{}c@{}}vc-relax \\ PBE0 \\ 0 K \\{(}eV{)}\end{tabular}} & \multicolumn{1}{c}{\begin{tabular}[c]{@{}c@{}}vc-relax \\ PBE0+SOC \\ 0 K \\{(}eV{)}\end{tabular}} & \multicolumn{1}{|c}{\begin{tabular}[c]{@{}c@{}} $\textit{n} \times \textit{n}\times \textit{n}$ \\k-point \\ grid  \end{tabular}} \\ 
     \hline
    \multicolumn{1}{c|}{\begin{tabular}[c]{@{}c@{}}12-atom\end{tabular}}& \multicolumn{1}{c}{\begin{tabular}[c]{@{}c@{}}3.62 $\pm$ 0.21\\-\\-\\-\\-\\-\end{tabular}}& \multicolumn{1}{c}{\begin{tabular}[c]{@{}c@{}}5.08 $\pm$ 0.21\\-\\-\\-\\-\\-\end{tabular}} &  \multicolumn{1}{c}{\begin{tabular}[c]{@{}c@{}} 3.68\\2.21\\1.48\\1.49\\1.49\\1.48\end{tabular}}&  \multicolumn{1}{c}{\begin{tabular}[c]{@{}c@{}} 3.72\\2.31\\1.97\\1.63\\1.56\\1.55\end{tabular}}& \multicolumn{1}{c}{\begin{tabular}[c]{@{}c@{}} 3.46\\1.78\\1.25\\0.49\\0.45\\0.45\end{tabular}}&  \multicolumn{1}{c}{\begin{tabular}[c]{@{}c@{}} 5.82\\3.90\\3.47\\3.44 \\3.38\\3.37\end{tabular}} &\multicolumn{1}{c}{\begin{tabular}[c]{@{}c@{}}5.54\\3.33\\2.72\\2.28\\2.25\\2.24\end{tabular}} & \multicolumn{1}{|c}{\begin{tabular}[c]{@{}c@{}} 1\\2\\4\\6\\8\\10\end{tabular}} \\
    \rowcolor{Gray}    
    \multicolumn{1}{c|}{\begin{tabular}[c]{@{}c@{}}96-atom\end{tabular}} & \multicolumn{1}{c}{\begin{tabular}[c]{@{}c@{}}2.16 $\pm$ 0.17\\-\\-\\-\\-\end{tabular}} & \multicolumn{1}{c}{\begin{tabular}[c]{@{}c@{}}2.80 $\pm$ 0.15 \\-\\-\\-\\-\end{tabular}} & \multicolumn{1}{c}{\begin{tabular}[c]{@{}c@{}}2.24\\ 1.58\\1.49\\1.48\\1.61$^\star$\end{tabular}} & \multicolumn{1}{c}{\begin{tabular}[c]{@{}c@{}}2.26\\ 1.69\\1.58\\1.61\\1.60$^\star$\end{tabular}} & \multicolumn{1}{c}{\begin{tabular}[c]{@{}c@{}}1.78\\0.60\\0.46\\0.54\\0.48$^\star$\end{tabular}} & \multicolumn{1}{c}{\begin{tabular}[c]{@{}c@{}}3.36\\2.84\\2.73\\2.75\\2.76$^\star$\end{tabular}}& \multicolumn{1}{c}{\begin{tabular}[c]{@{}c@{}}2.85\\(1.75)\\(1.61)\\(1.68)\\(1.64)$^\star$\end{tabular}}& \multicolumn{1}{|c}{\begin{tabular}[c]{@{}c@{}} 1 \\ 2 \\ 4 \\ 6 \\ 6 \end{tabular}} \\
    \multicolumn{1}{c|}{\begin{tabular}[c]{@{}c@{}}768-atom\end{tabular}} & \multicolumn{1}{c}{\begin{tabular}[c]{@{}c@{}}1.76 $\pm$ 0.04 \\ - \\ - \end{tabular}} & \multicolumn{1}{c}{\begin{tabular}[c]{@{}c@{}}2.28 $\pm$ 0.03 \\ - \\ - \end{tabular}} & \multicolumn{1}{c}{\begin{tabular}[c]{@{}c@{}}1.54 \\ 1.74$^\star$ \\ 1.48\end{tabular}} & \multicolumn{1}{c}{\begin{tabular}[c]{@{}c@{}}1.60 \\ 1.74$^\star$ \\ -\end{tabular}}  & \multicolumn{1}{c}{\begin{tabular}[c]{@{}c@{}}- \\ - \\-\end{tabular}}& \multicolumn{1}{c}{\begin{tabular}[c]{@{}c@{}}- \\ - \\ -\end{tabular}} & \multicolumn{1}{c}{\begin{tabular}[c]{@{}c@{}}- \\ - \\ -\end{tabular}} & \multicolumn{1}{|c}{\begin{tabular}[c]{@{}c@{}}  1 \\ 1 \\ 2\end{tabular}} \\
    \rowcolor{Gray}   
    \multicolumn{1}{c|}{\begin{tabular}[c]{@{}c@{}}2592-atom\end{tabular}} & 1.59 $\pm$ 0.07 & 2.17 $\pm$ 0.03& \multicolumn{1}{c}{\begin{tabular}[c]{@{}c@{}}1.51 \\ 1.50$^\star$ \end{tabular}} & \multicolumn{1}{c}{\begin{tabular}[c]{@{}c@{}}1.71 \\ 1.50$^\star$ \end{tabular}} & - & - & - & \multicolumn{1}{|c}{\begin{tabular}[c]{@{}c@{}}  1 \\ 1\end{tabular}} \\
    \multicolumn{1}{c|}{\begin{tabular}[c]{@{}c@{}}6144-atom\end{tabular}} & 1.47 $\pm$ 0.08& 2.09 $\pm$ 0.05 & \multicolumn{1}{c}{\begin{tabular}[c]{@{}c@{}}1.47 \\ 1.68$^\star$ \end{tabular}}  &  \multicolumn{1}{c}{\begin{tabular}[c]{@{}c@{}}1.67 \\ 1.69$^\star$ \end{tabular}} & - & -& - & \multicolumn{1}{|c}{\begin{tabular}[c]{@{}c@{}}  1 \\ 1\end{tabular}} \\
    \hline
  \end{tabular}}
  \begin{tablenotes}
    \item $^\star$ FA pseudo-randomly oriented 
  \end{tablenotes}
\end{table}

\begin{table}[ht!]
  \caption{Mean squared error (MSE) of the lattice vectors with respect to the perfect cubic FAPbI$_3$ $\alpha$-phase (column 1) and absolute amplitude of the octahedra tilting angle (column 2) at 0 K. For the structure optimized from all-aligned FA, the octahedra tilting amplitude refers to the global deformation of the octahedra, while for the structure optimized from pseudo-random FAs it refers to the average absolute value of the octahedra tilting angle (Supplementary Fig.3). FAPbI$_3$ dipole moment per stechiometric unit (ABX$_3$) for different simulation cells; the dipole moment was computed for at 0 K (column 3) and after equilibration as average along the AIMD trajectories performed at 300 K in NPT-F (columns 4-5). The initial configurations for the 0 K simulations were chosen with all FAs aligned except the one indicated with a star superscript.}
  \label{tbl:DistDip}
  \centering
  \begin{tabular}{c|ccccc}
    \hline
    \rowcolor{Gray}
     \begin{tabular}[c]{@{}c@{}}Simulation\\ cell \end{tabular} &  \multicolumn{1}{c}{\begin{tabular}[c]{@{}c@{}} MSE \\ 0 K \end{tabular}} & \multicolumn{1}{c}{\begin{tabular}[c]{@{}c@{}}  Octahedra \\ tilting amplitude \\ 0 K \\ {(}deg{)} \end{tabular}} & \multicolumn{1}{c}{\begin{tabular}[c]{@{}c@{}}Dipole \\ PBE \\ 0 K \\ {(}Debye / ABX$_3${)}\end{tabular}} &  \multicolumn{1}{c}{\begin{tabular}[c]{@{}c@{}}Dipole \\ PBE \\ 300 K\\ {(}Debye / ABX$_3${)}\end{tabular}} & \multicolumn{1}{c}{\begin{tabular}[c]{@{}c@{}}Dipole \\ PBE0 \\ 300 K \\ {(}Debye / ABX$_3${)}\end{tabular}} \\
     \hline
    \multicolumn{1}{c|}{\begin{tabular}[c]{@{}c@{}}12-atom \end{tabular}} & 0.08 & 1.37 & 13.17 & 2.32 $\pm$ 0.98 & 1.80 $\pm$ 0.70 \\
    \rowcolor{Gray}         
    \multicolumn{1}{c|}{\begin{tabular}[c]{@{}c@{}}96-atom \end{tabular}} & \multicolumn{1}{c}{\begin{tabular}[c]{@{}c@{}} 0.78 \\ 0.09$^\star$ \end{tabular}} & \multicolumn{1}{c}{\begin{tabular}[c]{@{}c@{}} 18.03 \\ 17.36$^\star$ \end{tabular}} & \multicolumn{1}{c}{\begin{tabular}[c]{@{}c@{}} 3.36 \\ 3.25$^\star$ \end{tabular}} & 2.24 $\pm$ 1.02 & 4.57 $\pm$ 0.71\\ 
    \multicolumn{1}{c|}{\begin{tabular}[c]{@{}c@{}}768-atom \end{tabular}} & \multicolumn{1}{c}{\begin{tabular}[c]{@{}c@{}} 0.30 \\ 4$\cdot 10^{-3\star}$ \end{tabular}} & \multicolumn{1}{c}{\begin{tabular}[c]{@{}c@{}} 5.65 \\ 6.46$^\star$  \end{tabular}} & \multicolumn{1}{c}{\begin{tabular}[c]{@{}c@{}}1.18 \\ 1.10$^\star$ \end{tabular}} & 0.92 $\pm$ 0.26 & 0.89 $\pm$ 0.25\\ 
    \rowcolor{Gray}   
    \multicolumn{1}{c|}{\begin{tabular}[c]{@{}c@{}}2592-atom \end{tabular}} & \multicolumn{1}{c}{\begin{tabular}[c]{@{}c@{}}  0.23 \\ 8$\cdot 10^{-4\star}$ \end{tabular}} & \multicolumn{1}{c}{\begin{tabular}[c]{@{}c@{}} 5.18 \\ 5.67$^\star$ \end{tabular}} & \multicolumn{1}{c}{\begin{tabular}[c]{@{}c@{}} 0.39 \\ 0.36$^\star$ \end{tabular}}  & 0.39 $\pm$ 0.12 & 0.31 $\pm$ 0.09  \\   
    \multicolumn{1}{c|}{\begin{tabular}[c]{@{}c@{}}6144-atom\end{tabular}} & \multicolumn{1}{c}{\begin{tabular}[c]{@{}c@{}}  0.25 \\ 1$\cdot 10^{-4\star}$ \end{tabular}} & \multicolumn{1}{c}{\begin{tabular}[c]{@{}c@{}} 4.11 \\ 5.52$^\star$ \end{tabular}} & \multicolumn{1}{c}{\begin{tabular}[c]{@{}c@{}} 0.13 \\ 0.18$^\star$ \end{tabular}} & 0.24 $\pm$ 0.07 & 0.24 $\pm$ 0.07\\
    \hline
  \end{tabular}
  \begin{tablenotes}
    \item $^\star$ FA pseudo-randomly oriented 
  \end{tablenotes}
\end{table}

\subsection*{Dipole moment}
The dipole moment was computed at 0 K and at 300 K (Tab.\ref{tbl:DistDip}). In principle, the dipole moment should be null, but it is not because of a fictitious residual dipole due to the system size. Since the dipole moment decreases significantly with increasing system size, it has to be related to a long-range collective interaction such as FA rotation and/or octahedra orientation. Due to its conformation, FA has a non-zero dipole moment, while the octahedra orientation might induce a significant dipole moment. Tab.\ref{tab:FA_orient} reports the relative dipole moment obtained by subtracting the contribution of the dipole moment of the FAs in the initial configuration from the total dipole moment of the system. The result is about the same for all configurations (Tab.\ref{tab:FA_orient}), and comparable with the values in Tab.\ref{tbl:DistDip}, concluding that the initial FA configuration does not affect the overall dipole moment of the cell. To estimate the actual contributions of FA and octahedra to the dipole moment, two NPT-F simulations were performed with FA or octahedra frozen. The octahedra contribution is the one that compensates for the dipole moment of the system, while FA does not (Supplementary Fig.6). In light of this result, one might also expect that a too high and uncompensated FA dipole moment could result in the phase transition of FAPbI$_3$ from $\alpha$ to $\delta$. The time correlation function associated to the dipole moment fluctuations shows that the correlation time ($\sim$ 30 fs) is not so affected by the system size (Supplementary Fig.8) as the absolute value of the dipole moment. Except for the 12-atom and 96-atom cells, which we have already realized are too small, there is no major change in dipole moment when using PBE0. Finally, the values of the dipole moment obtained with PBE and PBE0 are statistically equivalent for the 6144-atom cell (Tab.\ref{tbl:DistDip}). This indicates that charge fluctuations does not hinge on the charge localization of the functional. 

\begin{table}[ht!]
    \caption{Dipole moment, FA contribution to the dipole moment, and difference between the dipole moment and the FA contribution to the dipole moment. The values were computed for a 96-atom cell equilibrated snapshot starting from an initial configuration with different FA orientations (Supplementary Fig.7).}
    \centering
  \begin{tabular}{c|ccc}
    \hline
    \rowcolor{Gray}
     \begin{tabular}[c]{@{}c@{}}FA \\ orientation \end{tabular} &  \multicolumn{1}{c}{\begin{tabular}[c]{@{}c@{}} Dipole$_{\text{tot}}$ \\ {(}Debye / ABX$_3${)}\end{tabular}} &  \multicolumn{1}{c}{\begin{tabular}[c]{@{}c@{}} Dipole$_{\text{FA}}$ \\ {(}Debye / ABX$_3${)}\end{tabular}} & \multicolumn{1}{c}{\begin{tabular}[c]{@{}c@{}} Dipole$_{\text{tot}}$ - Dipole$_{\text{FA}}$ \\ {(}Debye / ABX$_3${)}\end{tabular}} \\
     \hline
        aligned & 0.82 & 0.30 & 1.02 \\
        \rowcolor{Gray}   
        random & 1.00 & 0.12 & 1.08 \\
        random\_best & 0.94 & 0.06 & 1.02 \\
        \rowcolor{Gray}   
        smart\_100 & 1.05 & 0.00 & 1.05 \\
        smart\_quasi & 1.16 & 0.00 & 1.16 \\
    \end{tabular}
    \label{tab:FA_orient}
\end{table}

\section*{Discussion}
In conclusion, by means of large-scale first-principle calculations and ab initio molecular dynamics simulations at 300 K, we demonstrated that in order to get an accurate description of the structural and electronic properties of FAPbI$_3$ $\alpha$-phase the size of the simulated system needs to approach the nano-scale. In particular, we showed that three conditions have to be met simultaneously, namely a proper description for the band gap, the minimization of structural distortions, and the zeroing out of the dipole moment. For first-principle calculations it is essential to start from an initial configuration where the FAs are pseudo-randomly oriented by preserving the 3-fold symmetry and minimizing the dipole moment. At 300 K, because of the kinetics, the initial configuration of the FAs is not stringent and, from a 2592-atom cell upwards, the PBE approximation is already able to describe the electronic band gap of the $\alpha$-FAPbI$_3$. For the 6144-atom cell we have computed a band gap of $1.47 \pm 0.08$ eV which is in excellent agreement with the experimental values of 1.45-1.51 eV reported in literature (highlighting that PBE0 and SOC corrections only cancel out for this system size range). The same cell minimizes structural distortions with respect to the perfect $\alpha$-FAPbI$_3$ structure and has the lowest dipole moment among all the systems studied. 
A significant correlation was discovered between PbI$_6$ octahedra tilting, band gap oscillations, and dipole moment. In particular, the dipole moment goes to zero only if the system size is big enough to properly relax the tilting pattern of the octahedra. Overall, an adequate size of the system (at least 6144-atom cell) is needed to correctly describe its physics, as we have demonstrated with the identification of band gap domains related to a correct description of the octahedra tilting. Our work provides a detailed insight into the connection between structural and electronic properties of $\alpha$-FAPbI$_3$ - and MHPs in general - making an important contribution to the field of ab initio simulations dedicated to understanding fundamental physical principles, such as hole-electron transport, which is of paramount importance in the development of increasingly high-performance PSC devices.

\section*{Methods}
\subsection*{First-principle calculations}
DFT simulations at 0 K were performed with Quantum ESPRESSO (QE)\cite{giannozzi_quantum_2009} suite of codes. All the calculations were run with DOJO fully relativistic norm-conserving PBE pseudopotentials\cite{van_setten_pseudodojo_2018,hamann_optimized_2013} and well-converged basis sets corresponding to an energy cutoff of 150 Ry for the wave functions and 600 Ry for the charge density. Different k-point Monkhorst-Pack grids\cite{monkhorst_special_1976} were used, all centered on $\Gamma$-point. Semiempirical corrections accounting for the van der Waals interactions were included with the DFT-D3 approach.\cite{grimme_density_2011} Different simulation cells were used, starting from a $1\times1\times1$ $\alpha$-FAPbI$_3$ (12-atom) up to a $4\times4\times4$ (768-atom). The electronic structure of fully relaxed structures (vc-relax) was also computed including spin-orbit coupling (SOC) and PBE0. The supercell distortion with respect to the perfect cubic $\alpha$-FAPbI$_3$ has been estimated by the mean squared error (MSE) between the lattice vectors of the two systems. Because CP2K\cite{kuhne_cp2k_2020} allows to run also DFT at 0 K, $\Gamma$-point simulations were performed with both the QE and CP2K software, obtaining equal band gap values to three decimal places, which means that the results achieved with the two software are comparable.

\subsection*{Ab initio molecular dynamics}
AIMD simulations were run in the DFT framework as implemented in the CP2K software. The PBE and PBE0 functional and the D3 dispersion correction\cite{grimme_density_2011} were adopted together with Goedecker-Teter-Hutter pseudopotentials\cite{perdew_restoring_2008} and a polarized double-$\zeta$ Gaussian basis set (DZVP)\cite{vandevondele_gaussian_2007} for valence electrons. The energy cut off for the expansion of the electron density was set to 400 Ry. Simulations were run with a time step of 0.5 fs in the NPT flexible ensemble using Born-Oppenheimer dynamics for 7-12 ps (PBE) and 2-5 ps (PBE0), while the temperature was controlled by the Bussi thermostat\cite{bussi_canonical_2007} and the pressure by the Martyna barostat.\cite{martyna_constant_1994} Different simulation cells were used, starting from a $2\times2\times2$ $\alpha$-phase FAPbI$_3$ (96-atom) up to a $8\times8\times8$ (6144-atom). The band gap was computed in temperature  as an average of different band gaps calculated from the projected density of states (PDOS) on several AIMD snapshots after the system equilibrated ($\sim2$ ps). The spatial variation of the band gap within a supercell was calculated by grouping the PDOS of the atoms of interest. Real space position of top of the valence and bottom of the conduction bands in FAPbI$_3$ were computed after quenching an equilibrated AIMD snapshot to 0 K. Different initial FA orientation were tested - completely ordered, random oriented, smart oriented (total FA dipole equal to zero) - to avoid any bias on the simulations. The dipole moment was calculated at the quantum level as in the CP2K framework,  while the contribution of FAs to the dipole in the initial configurations was estimated classically, assigning a +1 charge to each FA.

\subsection*{Time correlation function analysis}
The rotational dynamics of FA and PbI$_6$ octahedra were characterized by the correlation function
\begin{equation}
    C_{rot}(t)= \frac{\left\langle \vec{\mu}(t)\cdot\vec{\mu}(0) \right\rangle}{|\vec{\mu}(0)|},
\end{equation}
where $\vec{\mu}(t)$ is the C-H(N-N) vector for FA or the octahedra tilting function appropriate for the PbI$_6$ tilting.

The timescale oscillations for the band gap were quantified by the correlation function
\begin{equation}
    C_{gap}(t)= \frac{\left\langle \Delta\epsilon_{cv}(t)\cdot\Delta\epsilon_{cv}(0) \right\rangle}{|\Delta\epsilon_{cv}(0)|},
\end{equation}
where $\Delta\epsilon_{cv}(t)$ is the difference between the eigenvalues of the bottom of the conduction band and the top of the valence band. The same time correlation function has been used to compute the correlations of the dipole moment fluctuations, where the quantity correlated in time was the value of the dipole moment.

\clearpage
\bibliography{Pero}

\section*{Acknowledgements}
U.R. acknowledges the Swiss National Foundation (grant N. 200020\_219440) and computational resources from the Swiss National Computing Centre CSCS (project s1151).
 V.C. acknowledges computational resources from the Swiss National Computing Centre CSCS (project s1253).

\section*{Author contributions statement}
V.C. and L.A. conceived the idea. V.C, L.A, and V.S. performed the theoretical simulations and wrote the paper under the supervision of U.R.. N.L. and A.V. contributed to the discussion and writing of the paper.

\section*{Competing interests}
The authors declare no competing interests.

\section*{Additional information}
The online version contains supplementary material available at \\
Data and analysis scripts are available on Zenodo at 

\end{document}


\maketitle

\clearpage
\subsection*{FA 3-fold symmetry}
\begin{figure}
    \centering
    \includegraphics[width=1\linewidth]{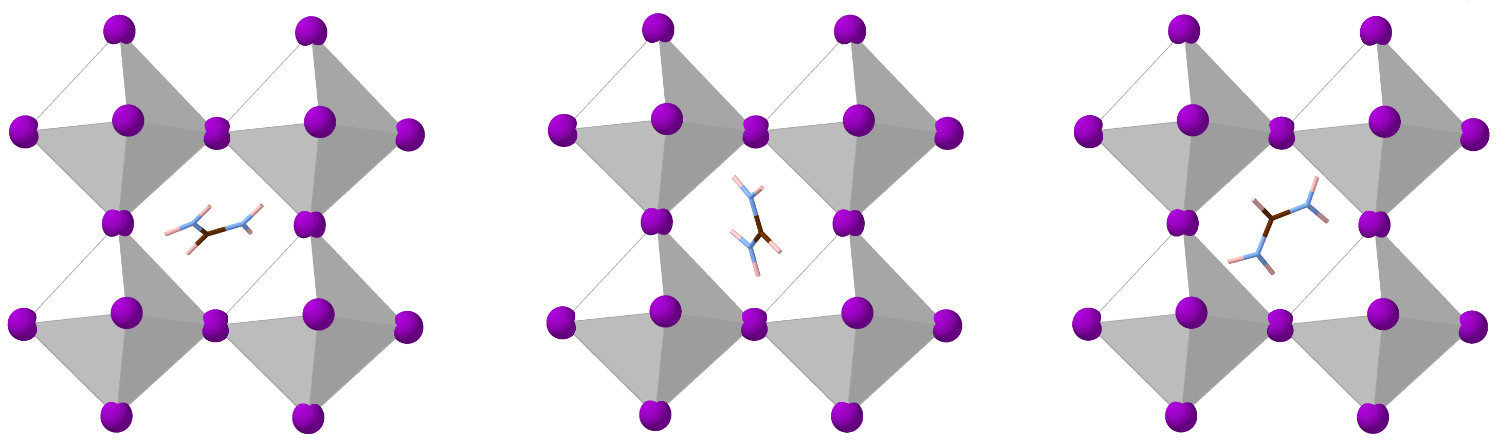}
    \caption{Orientations of the FA molecule satisfy the 3-fold symmetry within the PbI$_6$ cage. I, Pb, C, N, and H are shown in violet, white octahedra, brown,  lightblue, and pink, respectively.}
    \label{fig_S:3-fold}
\end{figure}

\clearpage
\subsection*{Octahedra tilting}

\begin{figure}
  \includegraphics[scale=1.3]{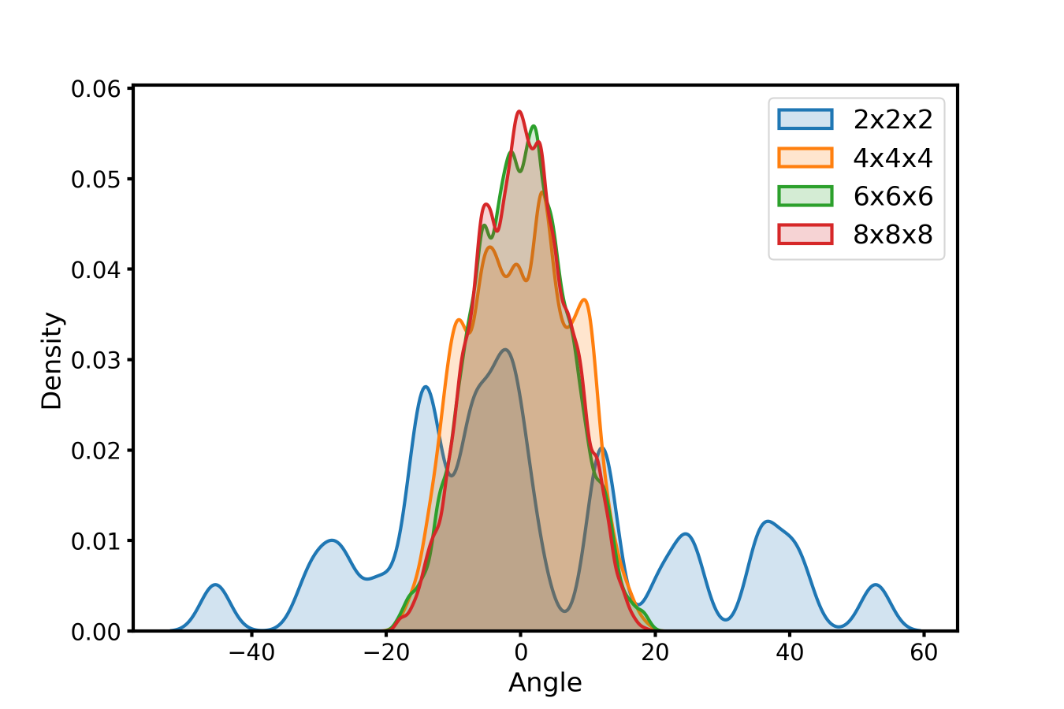}
  \caption{Distributions of the octahedra tilting angles for the different supercells optimized from a pseudo-random FA configuration.}
  \label{fgr_S:OctTiltDist}
\end{figure}

\begin{figure}
  \includegraphics[scale=0.93]{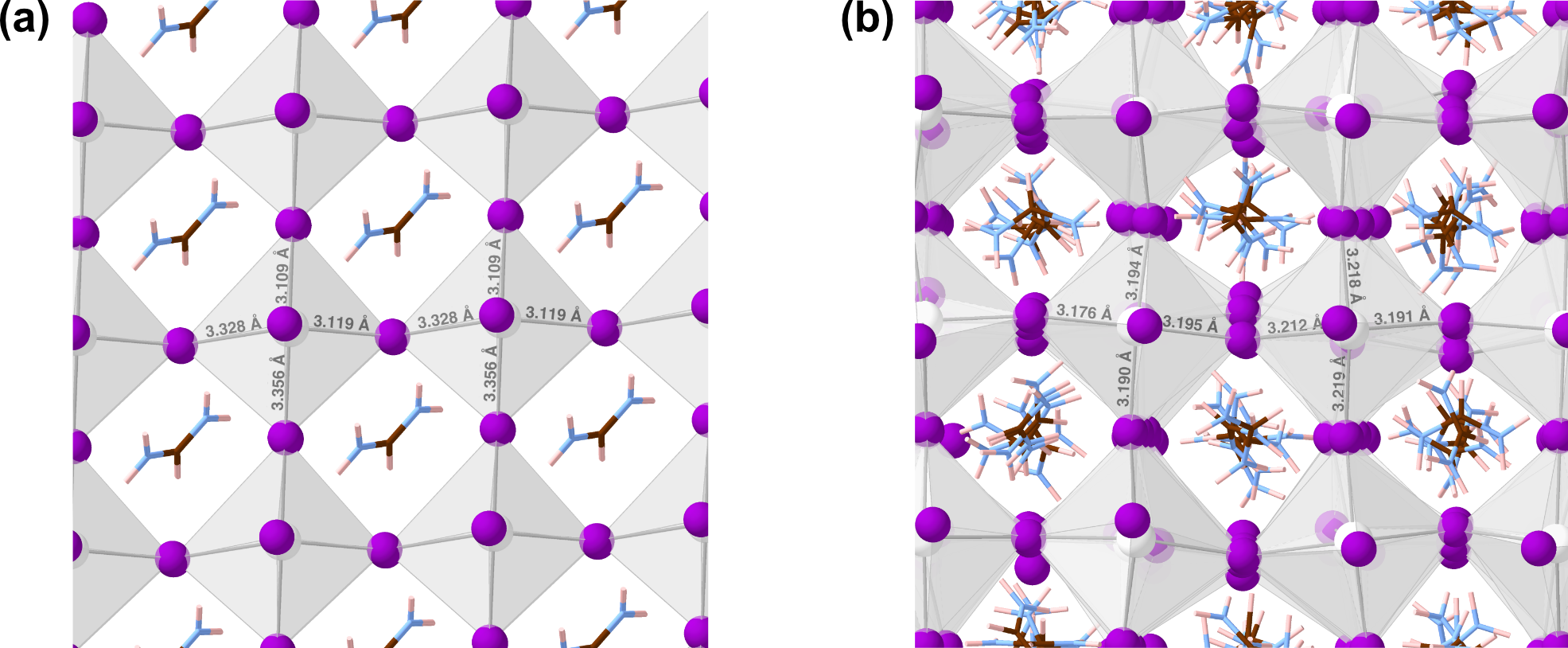}
  \caption{Zoom of the final FAPbI$_3$ structures starting from all-aligned (a) and pseudo-random (b) FA configurations after vc-relax. The Pb-I distances for two adjacent octahedra are highlighted. The typical tilting pattern of the octahedra can be recognised in the pseudo-random case, while the fully aligned case shows a non-physical collective I-shift. The colour code is the same as in Fig.\ref{fig_S:3-fold}.}
  \label{fgr_S:OctTilt}
\end{figure}

\clearpage
\subsection*{Band gap}


\begin{figure}
  \includegraphics[scale=1]{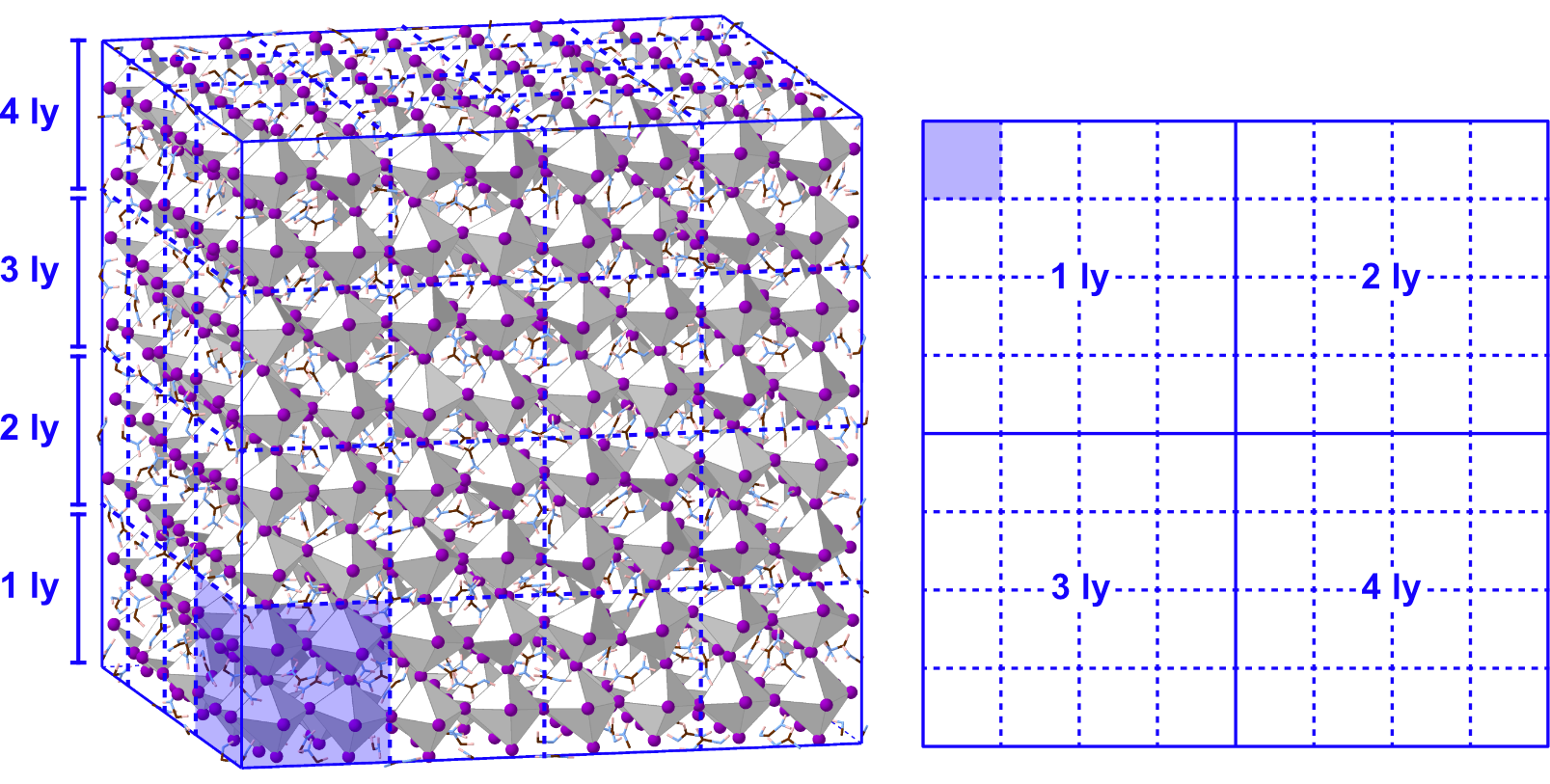}
  \caption{Band gap mapping of the 6144-cell from 3D to 2D. 64 96-cells were identified in the 6144-cell, as shown by the blue grid superimposed on the 3D model in the left panel, and rearranged into a 2D map as shown in the right panel. The area shaded in blue highlights a single 96-cell.  The colour code is the same as in Fig.\ref{fig_S:3-fold}.}
  \label{fgr_S:3Dto2D}
\end{figure}

\begin{figure}
  \includegraphics[scale=1.2]{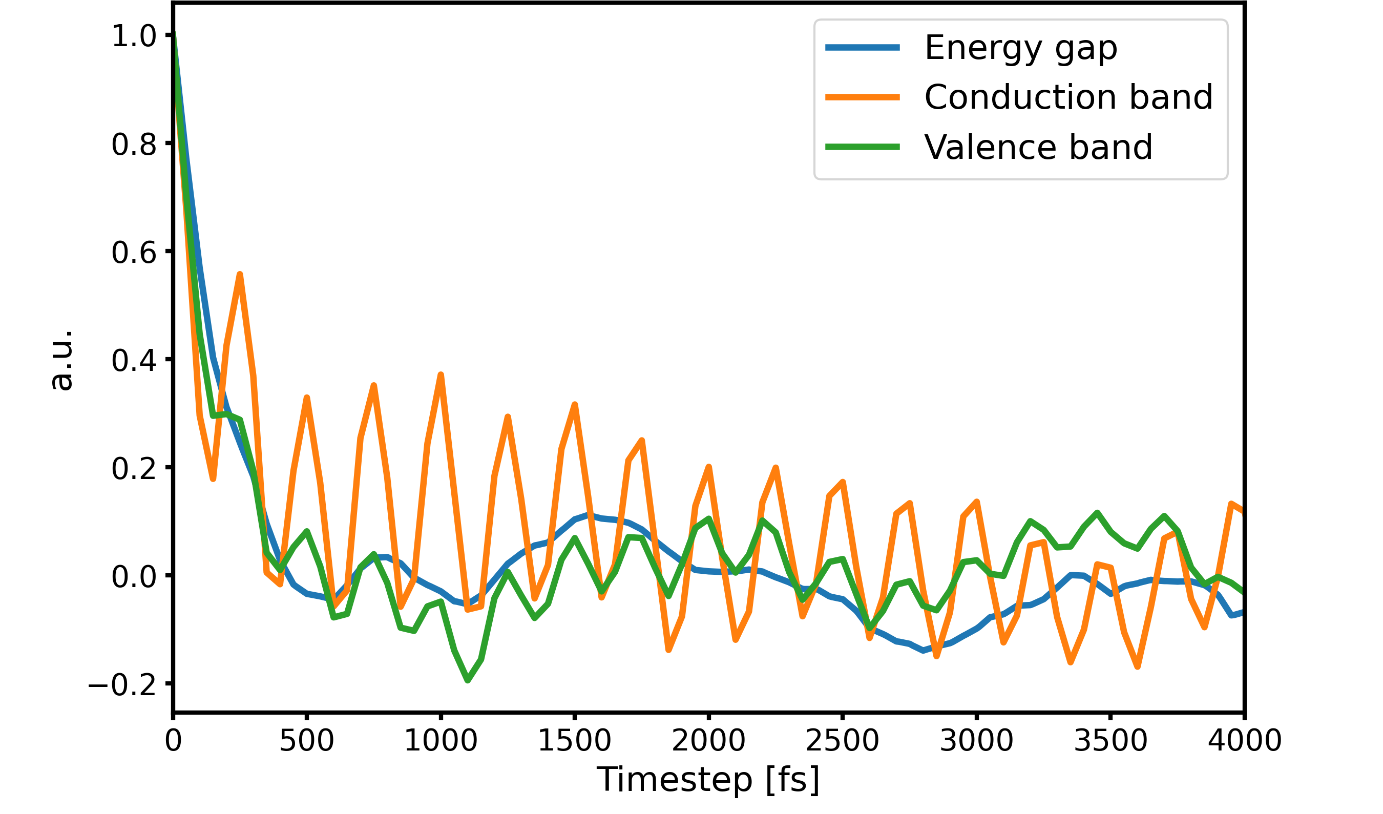}
  \caption{Time correlation function characterizing the band gap oscillations (blue curve), the bottom of the conduction band eigenvalue (orange curve), and the top of the valence band eigenvalue (green curve). The analysis was done on the 2592-atom one.}
  \label{fgr_S:CorrFuncBG}
\end{figure}

\clearpage
\subsection*{Dipole moment}

\begin{figure}
  \includegraphics[scale=1.45]{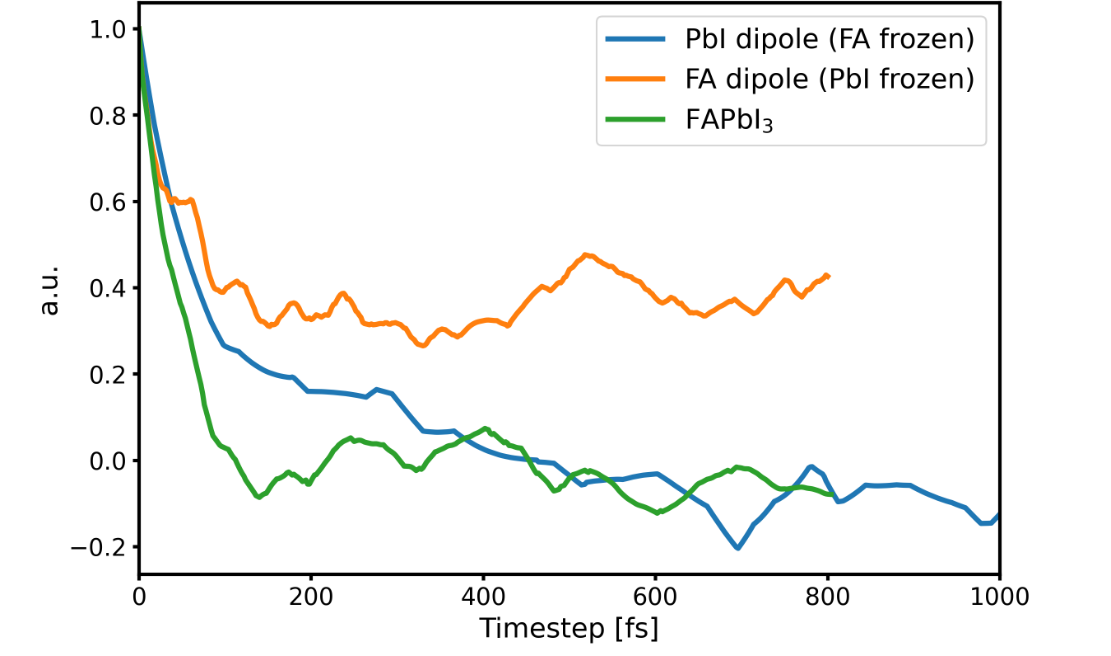}
  \caption{Time correlation function characterizing the dipole moment oscillations of the PbI$_6$ (orange curve), the FAs (orange curve), and the total one (green curve) for the 768-atom cell. Two NPT-F simulations were performed with FA or PbI$_6$ frozen to get the two separate contributions - PbI$_6$ and FA - to the dipole moment. FA alone cannot compensate the dipole moment of the lattice, but the PbI$_6$ fluctuations can (correlation goes to 0).}
  \label{fgr_S:Dipole}
\end{figure}

\begin{figure}
  \includegraphics[scale=1.1]{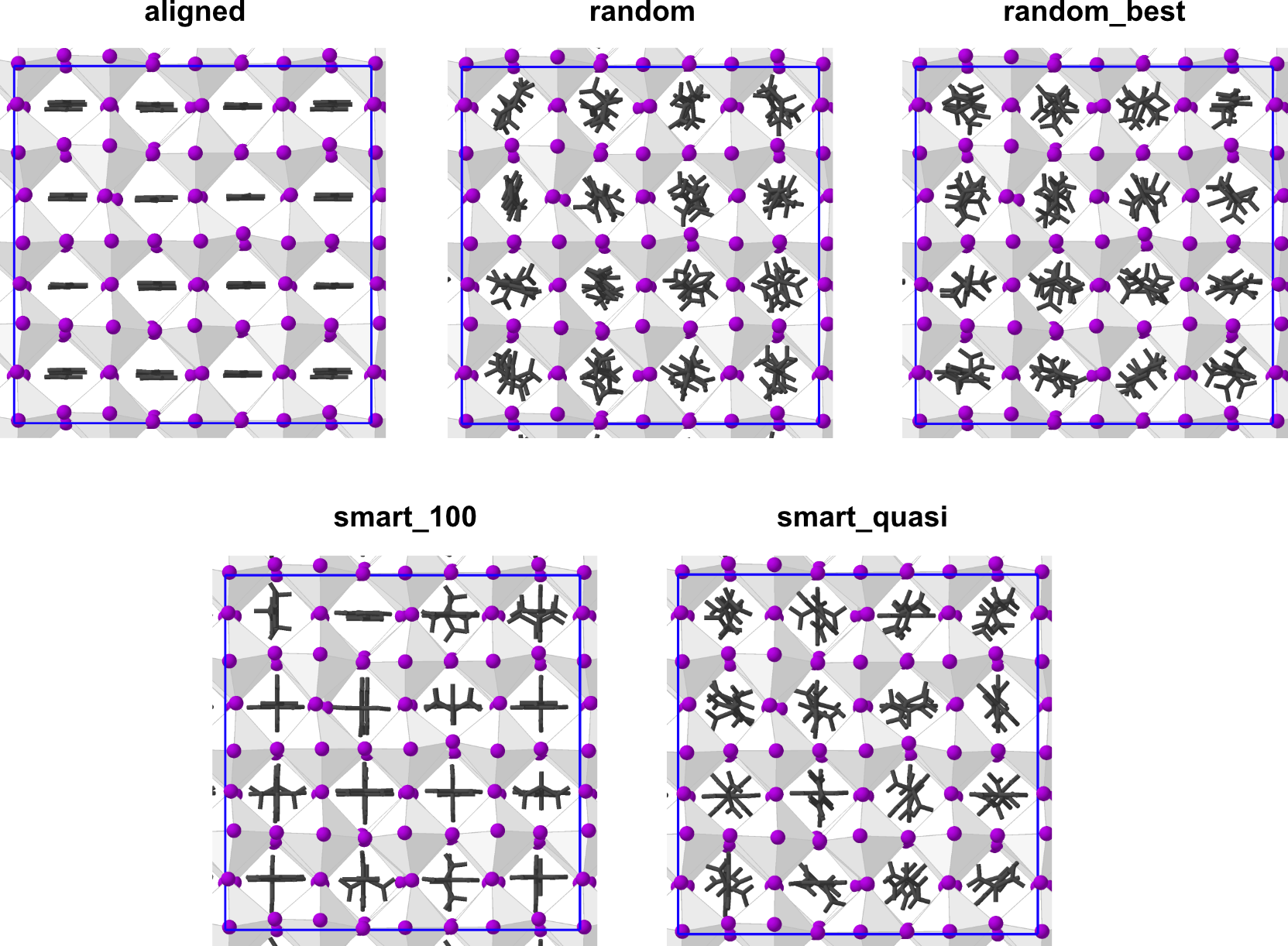}
  \caption{Initial configurations of the 768-atom cell with different FA orientations. FA molecules are shown in dark grey to facilitate the visualization. I and Pb are shown in violet and white octahedra, respectively.}
  \label{fgr_S:FAconf}
\end{figure}

Fig.\ref{fgr_S:FAconf} shows the different starting configurations of FAs for the 768-atom cell. In the "aligned" configuration all the FAs are aligned in the same direction, in the "random" configuration they are oriented randomly, in the "random\_best" configuration they are oriented randomly so as to minimize the dipole moment, in the "smart\_100" configuration they are oriented along the 100, 010, 001 directions minimizing the dipole moment, in the "smart\_quasi" configuration they are pseudo-randomly oriented preserving the 3-fold symmetry (Supplementary Fig.1) and minimizing the dipole moment.

\begin{figure}
  \includegraphics[scale=1.2]{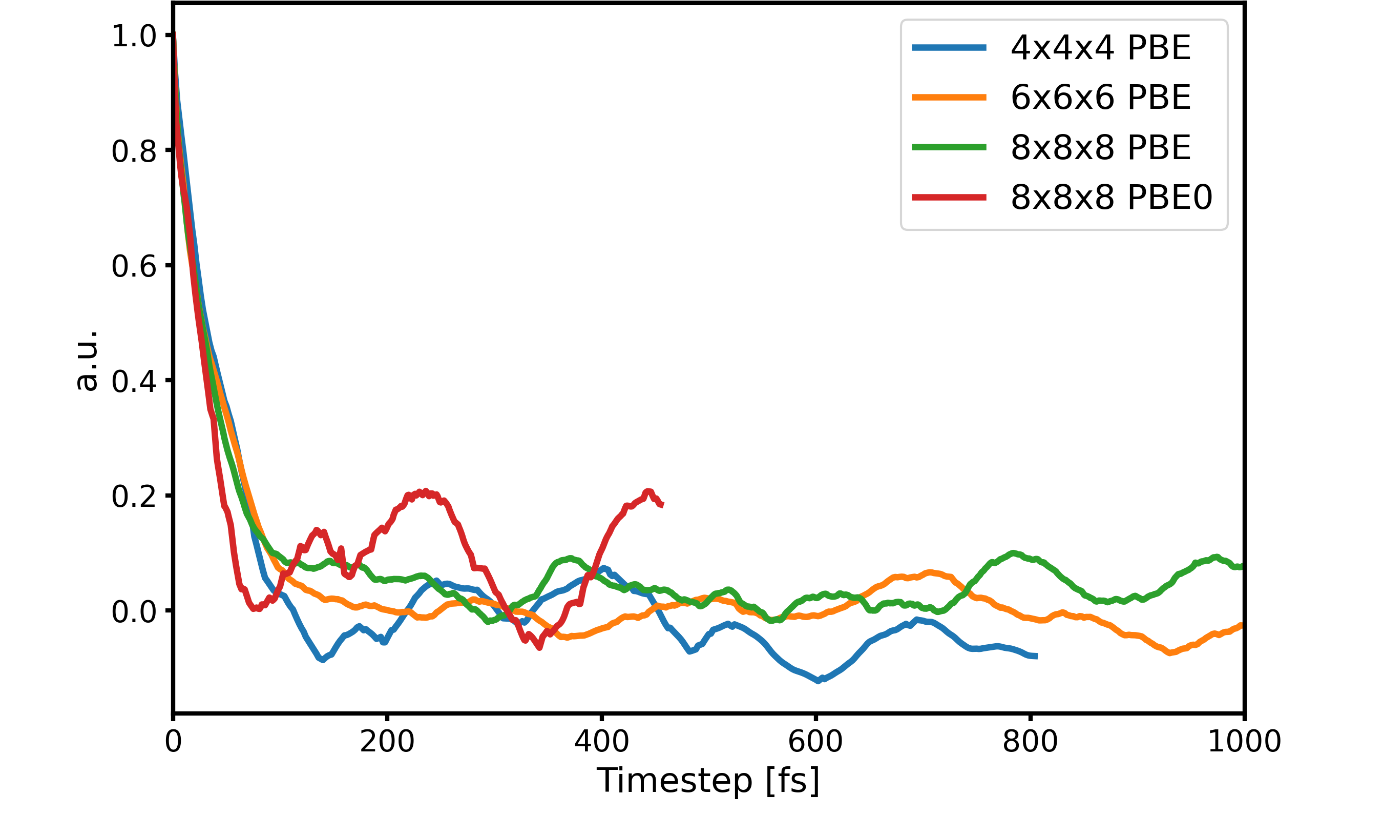}
  \caption{Time correlation function characterizing the dipole moment oscillations for different supercell size. The 6144-atom cell results are reported with both PBE and PBE0 level of theory.}
  \label{fgr_S:CorrFuncDip}
\end{figure}